\documentclass[reprint,aps,prd,amsmath,showpacs,aps]{revtex4-1}
\usepackage{graphicx}
\usepackage[linkcolor=blue,citecolor=blue]{hyperref}
\usepackage{bm}
\usepackage{float}
\usepackage{amssymb}
\usepackage{latexsym}
\usepackage{scrextend}
\usepackage{cleveref}
\hypersetup{colorlinks=true}

\def\<#1>{{\small #1}}
\begin{document}
\title{Applying Bayesian Inference to Galileon Solutions of the Muon Problem}

\author{Henry Lamm}
\email{hlammiv@asu.edu}

\affiliation{Physics Department, Arizona State University, Tempe, Arizona 
85287, USA}
\date{\today}

\begin{abstract}
We derive corrections to atomic energy levels from disformal couplings in Galileon theories.  Through Bayesian inference, we constrain the cutoff radii and Galileon scale via these corrections.  To connect different atomic systems, we assume the various cutoff radii related by a one-parameter family of solutions.  This introduces a new parameter $\alpha$ which is also constrained.  In this model, we predict shifts to muonic helium of $\delta E_{He^3}=1.97^{+9.28}_{-1.87}$ meV and $\delta E_{He^4}=1.69^{+9.25}_{-1.61}$ meV as well as for true muonium, $\delta E_{TM}=0.06^{+0.46}_{-0.05}$ meV.
\end{abstract}

\maketitle
\section{Introduction}
\label{sec:1}
Measurements in muon physics\cite{PhysRevD.73.072003,Antognini:1900ns,Aaij:2014ora,Pohl1:2016xoo} have shown discrepancies with
theoretical calculations.  This ``muon problem'' could signal  \textit{lepton universality} violation from beyond
standard model physics.  A stronger muon coupling to new physics is sensible from effective field theory (EFT).  Suppose the EFT has a cutoff scale $\Lambda$.    Then, observables should scale as powers of $m_l/\Lambda$.  This is analogous to the enhancement of weak interactions in muonic systems~\cite{PhysRevD.91.073008}.

Disformal scalar couplings can arise in Galileon theories currently being
investigated in modified gravity scenarios~\cite{Nicolis:2008in,Brax:2014vva}.  
The disformal coupling to matter allows for quantum loop corrections to atomic energy levels.  This opens up the tantalizing possibility that gravitational effects resolve the radii
discrepancies~\cite{Antognini:1900ns,Pohl1:2016xoo}.  It is necessary to include chameleon interactions to avoid constraints from astrophysical and colliders~\cite{Brax:2014zba}.  These interactions, as will be discussed below, introduce a mechanism for regularizing the divergence and explaining the origin of the Galileon radius. 

Due to the highly singular nature of the disformal 
scalar interaction, a particle-dependent cutoff radius $r_i$ for the Galileon interaction had to be introduced to render the $2s-2p$ Lamb shift finite.  Brax and Burrage assumed $r_i$ was equal to the 
particle charge radius $\sqrt{\left<r^2_{ch}\right>_i}$~\cite{Brax:2014zba}, but only considered bound states with nuclei.  In~\cite{Lamm:2015gka}, this assumption was applied to purely leptonic bound states (e.g. $e^+e^-$, $e^-\mu^+$).  The leptonic $r_i$ consistent with the muonic hydrogen discrepancy was found to be experimentally ruled out.  Therefore $r_i=\sqrt{\left<r^2_{ch}\right>_i}$ is inconsistent with the data.  

Removing this constraint, the relation between $r_i$ of different particles is must be specified some other way. The nonperturbative nature of the Galileon field makes computing $r_i$ from first principles difficult.  In this work, we instead introduce a phenomenological one-parameter relationship between $r_i$ of different particles.  
In~\cite{Lamm:2015gka}, it was seen that using the Lamb shift of multiple atoms is unable to break the degeneracy between $r_i$ and $M$ in parameter space.  To resolve this issue we  compute the Galileon correction to the $1s$ Lamb shift, $1s-2s$ interval, and the circular transitions between states $n\leq 5$.  These new constraints are found to partially break the degeneracy in regions of parameter space where sufficiently strong experimental bounds exist.

We begin in Sec.~\ref{sec:gal} with a short review of how disformal couplings arise and where the leading-order corrections to the transitions are found.  Sec.~\ref{sec:par} is devoted to introducing and motivating the model for $r_i$ used in this paper.  Following this is a short discussion of the transitions used in our study in Sec.~\ref{sec:tran}.  In Sec.~\ref{sec:anal} are found the results from considering all the experimental values in a Bayesian analysis.  Using the results, Sec.~\ref{sec:hel} presents prediction for the Galileon correction to muonic helium.  We conclude in ~\ref{sec:sum} with a short discussion of future work.

\section{Corrections from Galileons}
\label{sec:gal}
Bekenstein has shown that the most general metric formed from only $g_{\mu\nu}$ 
and a scalar field $\phi$ respecting causality and weak equivalence 
is~\cite{Bekenstein:1992pj}:
\begin{equation} 
\tilde{g}_{\mu\nu}=A(\phi,X)g_{\mu\nu}+B(\phi,X)\partial_\mu\phi\partial_\nu\phi
,
\end{equation}
where $X=\frac{1}{2}g_{\mu\nu}\partial_\mu\phi\partial_\nu\phi$.  The first term 
leads to conformal scalars, whose couplings to matter are heavily constrained by 
various fifth-force experiments.  For us, only the second term, which yields the 
disformal coupling, matters.  This Lagrangian interaction is
\begin{equation}
 \mathcal{L}_{dis}=\frac{B(\phi,X)}{2}\partial_\mu\phi\partial_\nu\phi 
T^{\mu\nu}_{J},
\end{equation}
where $T^{\mu\nu}_{J}$ is the energy-momentum tensor of matter given in the 
Jordan frame.  
 
The leading disformal coupling in nonrelativistic systems is a one-loop quantum 
effect that results in a correction to the energy level of an atomic system 
given by~\cite{Kugo:1999mf,Kaloper:2003yf,Brax:2014vva}:
\begin{equation}
\label{eq:de1}
 \delta E=-\frac{3m_im_j}{32\pi^3M^8}\bigg< E\bigg|\frac{1}{r^7}\bigg|E \bigg>,
\end{equation}

where $m_i\geq m_j$ are the masses of the constituent particles and $M$ is the 
Galileon coupling scale.  

From this we can derive the correction to each energy 
level.  For the $n=1,2$ states, the correction diverges like $1/r^4$ and 
therefore must be cutoff at some radius for each mass $m_i$, $r_i$.  For the 
$n=3$, the correction has a milder singularity of $\log(r)$.  For states $n\geq 
4$, the correction is finite in the limit of $r\rightarrow 0$, and therefore 
these higher transitions can give limits on $M$ that are less dependent on 
$r_i$.  

Our results for the corrections to the transition energies are 
listed in in Table~\ref{tab:eqs}.  We note that these are the exact relations 
obtained from using the full hydrogenic wave functions, in contrast to previous 
works\cite{Brax:2014vva,Brax:2014zba,Lamm:2015gka}.  Using the full wave 
functions was found to be necessary when rederiving the $2s-2p$ Lamb shift 
correction.  There, the next-to-leading-order term in the $2s$ state 
is larger than the leading-order $2p$ term, and therefore the energy 
correction used in~\cite{Brax:2014vva,Brax:2014zba,Lamm:2015gka} is 
inconsistent.  Due to the small size of these corrections in comparison to the leading-order $2s$ term, previous results are unaffected except for very large $r_i$.

\begin{table*}
 \caption{\label{tab:eqs}$\delta E_n=\kappa_n(x) 
F_n(x)$,$\eta=\frac{m_im_j}{\pi^3M^8a^7}$,$x=r_i/a$,where $a=(Z\alpha m_r)^{-1}$ 
is the Bohr radius of the system, $m_r$ is the reduced mass, and we have defined 
a function,where $\text{Ei}(x)$ is the exponential integral function.}
 \begin{center}
 \begin{tabular}{l c c}
 \hline\hline
  $n$&$\kappa_n(x)$&$F_{n}(x)$\\
  \hline
  1s Lamb& 
$-\frac{\eta}{2^5x^4}$&$e^{-2x}(3-2x+2x^2-4x^3)-8x^4\text{Ei}\left(-2x\right)$\\
 
1s-2s&$\frac{\eta}{2^8x^4}$&$8e^{-2x}\left(3-2x+2x^2-4x^3\right)-e^{-x}
\left(3-5x+4x^2-4x^3\right)-4x^4\left[16\text{Ei}(-2x)+\text{Ei}(-x)\right]$\\
   2s-2p 
Lamb&$\frac{\eta}{2^9x^4}$&$e^{-2x}(6-10x+7x^2-7x^3)-7x^4\text{Ei}
\left(-x\right)$\\
  
2p-1s&$\frac{\eta}{2^9x^4}$&$2^4e^{-2x}(3-2x+2x^2-4x^3)-x^2e^{-x}(1-x)-x^4\left[
2^7\text{Ei}(-2x)-\text{Ei}(-x)\right]$\\
  3d-2p&$-\frac{\eta}{2^8x^4}$ & 
$e^{-x}(-3+5x-4x^2+4x^3)+4x^4\left[\text{Ei}(-x)-\frac{2^4}{3^85}\text{Ei}
\left(-\frac{2}{3}x\right)\right]$\\
   
4f-3d&$-\frac{\eta}{2^{20}3^15^17^1}$&$e^{-\frac{x}{2}}(2+x)+\frac{2^{18}7}{3^7}
\text{Ei}\left(-\frac{2}{3}x\right)$\\
5g-4f&$\frac{\eta}{2^{20}3^15^{1}7^1}$&$e^{-\frac{x}{2}}(2+x)-\frac{2^{15}}{
3^25^{11}}e^{-\frac{2x}{5}}(3^15^3+150x+30x^2+4x^3)$\\
  \hline\hline
 \end{tabular}
\end{center}
\end{table*}

\section{Parametrizing with $r_G$}
\label{sec:par}

As seen in \cite{Lamm:2015gka}, the combination of multiple bound states can 
restrict the $(r_i,M)$ parameter space if the relationship 
between the Galileon radii is known.  The nonperturbative nature of the Galileon
field with chameleon traits makes computing $r_i$ from first principles at least
as difficult as computing the charge radii~\cite{Green:2014xba,Capitani:2015sba}, and requires choosing a particular chameleon field interaction which introduces model dependence.  
For this work, we instead develop a phenomenological relationship between $r_i$ 
of different particles motivated by general features of chameleon models and 
field distributions. 

Following~\cite{Lamm:2015gka}, we take the view that the Galileon radii should be interpreted as other radii, as an expectation value of an underlying distribution. Formally the charge  (where we mean charge in the general sense, e.g. electric charge, weak interaction, matter density) radius of a particle is defined via the associated form factor,
\begin{align}
 G_i(q^2)=&\int\mathrm{d}^3xe^{i\bm{q}\cdot \bm{x}}\rho(\bm{x})\nonumber\\&=\int\mathrm{d}^3x\left(1+i\bm{q}\cdot\bm{x}+\frac{(\bm{q}\cdot\bm{x})^2}{2}+\cdots\right)\rho(\bm{x})\nonumber\\&=Q_{i,\rm tot}-\frac{1}{6}|\bm{q}|^2\langle r^2\rangle+\cdots,
\end{align}
where $G_i$ is the form factor, $\rho(\bm{x})$ is the charge density, and $Q_{\rm tot}$ is the total charge of the particle.  The standard definition of $\langle r^2\rangle$ is then
\begin{equation}
\label{eq:rc}
 \langle r^2_i\rangle\equiv r_i^2=-6\frac{dG_i}{dq^2}\bigg|_{q^2=0}.
\end{equation}
By this definition, we see that $r_i$ is related to a Galileon density $\rho_{\rm G}(x)$ which measures the spatial distribution of matter coupling to the Galileons,
\begin{equation}
\label{eq:rr}
 r^2_i=\int\mathrm{d}r\mathrm{d}\theta\mathrm{d}\phi\sin(\theta) r^4\rho_{\rm G}(\bm{r}).
\end{equation}

In order to produce a viable phenomenological model of $r_i^2$, we therefore need an approximation for $\rho_{\rm dis}(\bm{r})$.  To do this, we first digress to discuss chameleon models.  Chameleon fields are scalar fields with density-dependent masses.  In cosmology and astrophysics, this feature is used avoid constraints on their production in the early Universe and star, while allowing them to be a dark energy candidate.  These fields are fully characterized by their mass and coupling constants.  One example of chameleons is the large curvature $f(R)$ model~\cite{Brax:2007ak,Dvali:2010jz}, which has a known function $m_\chi(\rho_m)=m_{\chi,0}\left(\rho_m/\rho_0\right)^{(n+2)/2}$
where $m_\chi$ is the Galileon mass, $\rho_0$ is the matter density of the Universe today, and $n$ is a model-dependent positive index.  

From this example, it is obvious to understand why stellar constraints can be avoided.  In vacuum, $m_\chi$ is nearly massless (present constraints are $\approx10^{-30}$ eV).  In the interior of a star, the matter density $\rho_m\approx10^{40}\rho_0$, implying that $m_\chi$ becomes large and suppresses the interaction.  The chameleon screening will have a more pronounced effect in leptons and nuclei where the density is even higher.  This should regularize the divergence in energy levels, rendering them finite, and justify the physical nature of $r_i^2$. 

With these properties in mind, we can propose a gross model for the Galileon radii.  Empirically, the density of nuclei $A>20$ is found to saturate at $\rho_{m,N}\approx100$ MeV/fm$^3$.  Neglecting shell effects, the matter radii can be related in the liquid-drop model by\cite{Gamow632}
\begin{equation}
 \rho_{m,N}=\frac{m_A}{\frac{4}{3}\pi r_A^3}.
\end{equation} 
implying $r_A\propto A^{\frac{1}{3}}$.  For $A<20$, the density is not saturated. We can estimate the density of the proton using its charge radius to be $\rho_{m,p}\approx300$ MeV/fm$^3$.  Conversely, if we estimate the proton radius from the saturation density, we obtain $r_{0}=1.2$ fm which is off by a factor of 1.4.  

If we can apply the liquid-drop model to the Galileon distribution, the chameleon screening effects should be the same and model independent for all particles and we would obtain
\begin{equation}
 \rho_{dis}(\bm{r})=\frac{C_G\rho_{m,N}}{4\pi}\times\theta(r_A-r)
\end{equation}
where $r_A=A^{\frac{1}{3}}r_0$, and $C_G$ is the correction factor from the chameleon interaction.  If we modify the standard definition of $A$ to be $A=m_A/m_p$, we can extend this definition to leptons as well.  With this, we can analytically evaluate Eq.(\ref{eq:rr}) to obtain $r_i=A^{\frac{1}{3}}r_G$ where $r_G=\sqrt{\frac{3}{5}}C_Gr_0$.  A more general model, which can be considered a perturbation from the uniform density model, is where the density now depends on radius
\begin{equation}
 \rho_{dis}(\bm{r})=\frac{\rho}{4\pi}\left[1+\left(\frac{r}{r'_A}\right)^n\right]\times\theta(r_A-r).
\end{equation}
The power $n$ in this model is determined by three things: the scaling of $\rho_m$ for a particle from the standard model interactions, $A$ to account for differences in particles, and the decoupling due to the model-dependent $m_{\chi}(\rho_m)$.  On general grounds, the competition between the first two mechanisms and the last will drive $|n|$ to smaller values and therefore a more uniform Galileon charge distribution.  Since $m_\chi$ becomes very large, this decoupling should not effect the matter distribution in the particle, similar to how the weak interaction has a negligible effect on nuclear structure.  In this model, the parameters $r_A=f(A)r_0,r'_A=g(A)r_0$ are two, as-yet undefined functions affected only by particle species.  Integrating, we find in this model that
\begin{equation}
\label{eq:rmod}
 r_i^2=\frac{3(n+3)(n+5\left[\left(\frac{f(A)}{g(A)}\right)^n+1\right])}{5(n+5)(n+3\left[\left(\frac{f(A)}{g(A)}\right)^n+1\right])}f(A)^2r_0^2.
\end{equation}
Assuming that $n$ is small and that $f(A),g(A)$ are slowly varying functions of $A$, the $A$ dependence of the numerator and denominator will be weak and tend to cancel.  Then, we can absorb the numerical factors and $r_0$ into $r_G$ and obtain $r_i\approx f(A)r_G$.  In this example, we see that essentially any function $f(A)$ can be specified for the relationship between radii and mass.

Motivated by these toy models, we propose a phenomenological one-parameter family of relations between the Galileon radii
\begin{equation}
\label{eq:rr}
 r_i=\left(\frac{m_i}{m_p}\right)^\alpha r_G,
\end{equation}
where $m_p$ is the proton mass, $r_G=r_p$ is the Galileon radius for the 
proton (which is unrelated to the charge radius, and to be determined), and 
$\alpha$ is a free parameter that will be fit by the data that relates different 
radii.   With Eq.~\ref{eq:rr}, corrections to transition energies from any bound state are determined by $(r_G,M,\alpha)$.  

This choice of parametrization can be further motivated by comparison to the charge radii.  In addition to the liquid-drop model discussed above,  power-law relations like Eq.~(\ref{eq:rr}) have found wide application.  Empirically fitting the $r_A$ for large elements, the relation $r_A=A^{0.294(1)}r_e$ is found to better account for the data~\cite{angeli2004consistent}.  In relating isotopic chains, $R_A=(A/A_0)^{1/5}R_0$ has been found to work well~\cite{Schiffer:1969cgk,Angeli:1977ncb}.  Accounting for the finite surface thickness of nuclei, the charge radii have been estimated using~\cite{elton1961nuclear}
\begin{equation}
 R_0=\left(r_0+\frac{r_1}{A^{2/3}}+\frac{r_2}{A^{4/3}}\right)A^{1/3},
\end{equation}
where a strong anticorrelation between $r_1$ and $r_2$ decreases the violation of the leading-order scaling with $A$.  With only the $\mu^-p$ and $\mu^-D$ results showing discrepancies, we believe that the model of Eq.~(\ref{eq:rr}) balances well the model dependence of using a more complex relation (with more free parameters) with the limited number of data points showing discrepancies.

While $\alpha=1/3,1/5,$ and $0.294(1)$ are all limiting cases of our model, there is a final case worth considering, that of $\alpha=0$.  This corresponds 
to the limit where all particles have the same $r_G$.  For this simplest case, we plot an example set of constraints in Fig.~\ref{fig:alp0}.

\section{Transitions}
\label{sec:tran}
Previous work on disformal scalars has focused almost exclusively on the discrepancies found in the $2s-2p$ Lamb shifts in muonic hydrogen and muonic deuterium.  In order to break the degeneracy between $r_G$ and $M$, it is useful to study the corrections to other atomic transitions where there is not an existing discrepancy.  We discuss the various experimental values that are used in our analysis in this section.  Throughout this work, we consider the energy difference $\Delta E_{exp-theor}$ which is the difference between the experimental and theoretical values.

The muonic hydrogen and muonic deuterium discrepancies\cite{Antognini:1900ns,Pohl1:2016xoo} we 
use are discrepancies between experimental values and theoretical calculations using the CODATA values of the charge radii\cite{Mohr:2015ccw} and are found in Tab.~\ref{tab:2sls}.   Along with these, we use the analogous constraints for muonium $(e^-\mu^+)$ and positronium 
$(e^-e^+)$\cite{Woodle:1990ky,Bhatt:1987zz,Erickson:1988zz,Lautrup:1971jf,PhysRevLett.71.2887,PhysRevA.60.2792}.  Since the Galileon correction is proportional to the mass of the two particles in the atom, leptonic system bounds are much weaker for $\alpha=0$ since $m_e,m_\mu\ll m_p, m_D$.  For $\alpha<1$, these limits move upward and become more constraining.  Leptonic systems then rule out small or negative $\alpha$ for all values of $r_G$, and $M$.  The muonium Lamb shift was only measured to 0.5\% in 1990, and a renewed experimental effort reducing this to match the 0.02\% theoretical uncertainty could significantly improve limits on new physics.  For positronium, the Lamb shift is also limited by experimental precision that is 2 orders of magnitude larger than the theoretical values. 

\begin{table}
 \caption{\label{tab:2sls}Difference between experiment and theory for $2s-2p$ 
Lamb shift in bound systems considered in this work.}
 \begin{center}
 \begin{tabular}{l c c}
 \hline\hline
  Atom&$\Delta E_{\rm exp-theor}~{\rm(meV)}$&Ref.\\
  \hline
  $\mu^- D$&0.438(59)&\cite{Pohl1:2016xoo}\\
  $\mu^- p$&0.329(47)&\cite{Antognini:1900ns}\\
  $e^- 
\mu^+$&$-2.3(9.6)\times10^{-5}$&\cite{Woodle:1990ky,Bhatt:1987zz,Erickson:1988zz,Lautrup:1971jf}\\
  $e^-e^+$&$4(695)\times10^{-8}$&\cite{PhysRevLett.71.2887,PhysRevA.60.2792}\\
  \hline\hline
 \end{tabular}
\end{center}
\end{table}

For muonium and positronium, it is also possible to use the $1s-2s$ interval to constrain the Galileon corrections.  The values adopted in this work are found in Table~\ref{tab:1s2s}.  While the $1s-2s$ intervals are also measured in hydrogen and deuterium, we neglect them due to their use in deriving the Rydberg constant and their theoretical uncertainty associated with QCD.  Compared to the $2s-2p$ Lamb shifts, the $1s-2s$ interval's experimental errors are only 1 order of magnitude larger than theory, so smaller gains are possible without theory improvements.

\begin{table}
 \caption{\label{tab:1s2s}Difference between experiment and theory for the 
$1s-2s$ interval in leptonic systems considered in this work}
 \begin{center}
 \begin{tabular}{l c c}
 \hline\hline
  Atom&$\Delta E_{\rm exp-theor}$ (meV)&Ref.\\
  \hline
$e^-\mu^+$	&$2.3(4.1)\times10^{-5}$	
&\cite{Meyer:1999cx,Karshenboim:1996bg,Pachucki:1996jw,Karshenboim:1997zu}\\
$e^-e^+$	&$2.4(3.5)\times10^{-5}$	
&\cite{PhysRevLett.70.1397,PhysRevA.60.2792}\\
  \hline\hline
 \end{tabular}
\end{center}
\end{table}

We also apply constraints from heavy hydrogenlike ions to restrict $\alpha>1$ since any limit in these systems becomes even more restrictive.   In the ions we investigated, the $1s$ Lamb shift has been measured to the 1\% level or less.  The results we utilize are found in Table~\ref{tab:1sls}.  The error in these results is dominated by experimental error, which is two orders of magnitude larger than the theoretical values, although ongoing work may improve these soon.   

\begin{table}
 \caption{\label{tab:1sls}Difference between experiment and theory for the $1s$ 
Lamb shift in heavy hydrogenlike ions considered in this work.}
 \begin{center}
 \begin{tabular}{l c c}
 \hline\hline
  Atom&$\Delta E_{\rm exp-theor}$ (eV)&Ref.\\
  \hline
 $e^-Pb^+$&15.4(22.0) &\cite{kraft2016precise}\\
  $e^-Au^+$&2.8(13.0) &\cite{kraft2016precise}\\
   $e^-Au^+$&-3.2(8.0) &\cite{Beyer1995}\\
    $e^-U^+$&-3.4(4.7) &\cite{PhysRevLett.94.223001,yerols}\\
  \hline\hline
 \end{tabular}
\end{center}
\end{table}

Higher $Z$ muonic atoms have been studied extensively, and their transitions can 
also be leveraged to constrain $r_G$ and $M$.  We note that the potential of 
Eq.~(\ref{eq:de1}) is not sensitive to spin, so the fine structure of the x-ray transitions are not effected. It would be interesting to compute Galileon corrections from the annihilation channel. This would open up both the fine structure and precision hyperfine splitting measurements to study. 

The most precisely measured transitions occur in $^{24}_{12}$Mg and $^{28}_{14}$Si, and 
these results have a large influence on the viable parameter space.  In the limit of $\alpha\rightarrow0$, they rule out Galileon corrections to muonic hydrogen and deuterium at a level far below those observed for $r_G<5\times 10^{-13}$ m for most $(r_G,M)$ and therefore drive $\alpha$ to positive values and $r_G$ to larger values (with the associated $M$ being driven lower).  The large set of muonic transitions used in this study are found in Table~\ref{tab:xray}.

\begin{table}
 \caption{\label{tab:xray}Difference between experiment and theory for muonic 
x-ray transitions considered in this work.}
 \begin{center}
 \begin{tabular}{l c c c}
 \hline\hline
  Element&Transition&$\Delta E_{\rm exp-theor}$ (eV)&Ref.\\
  \hline
$^{12}_6$C		&$2p_{3/2}-1s_{1/2}$	&$-3.8(1.6)$	
&\cite{Ruckstuhl:1984as}\\
$^{13}_6$C		&$2p_{3/2}-1s_{1/2}$	&$-1.8(7.2)$	
&\cite{DeBoer:1985rm}\\
$^{nat}_7$N		&$2p-1s$\footnote{\label{ufs}Unresolved fine structure}		
&$-2(11)$	&\cite{Dubler:1974lxk,Schaller:1980tit}\\
$^{nat}_8$O		&$2p-1s$\footref{ufs}		&$1(22)$	
&\cite{Dubler:1974lxk}\\
$^{24}_{12}$Mg		&$3d_{3/2}-2p_{1/2}$	&$0.7(1.1)$	
&\cite{Aas:1982bb,Aas:1982vs}\\
			&$3d_{5/2}-2p_{3/2}$	&$0.08(0.23)$	
&\cite{Beltrami:1985dc}\\
			&			&$-0.2(0.8)$	
&\cite{Aas:1982bb,Aas:1982vs}\\
$^{28}_{14}$Si		&$3d_{3/2}-2p_{1/2}$	&$0.6(2.0)$	
&\cite{Aas:1982bb,Aas:1982vs}\\
			&$3d_{5/2}-2p_{3/2}$	&$-0.18(0.33)$	
&\cite{Beltrami:1985dc}\\
			&			&$-0.4(1.2)$	
&\cite{Aas:1982bb,Aas:1982vs}\\
			&$4f_{5/2}-3d_{3/2}$	&$0.10(82)$	
&\cite{Beltrami:1984ss}\\
			&$4f_{7/2}-3d_{5/2}$	&$0.12(23)$	
&\cite{Beltrami:1984ss}\\
$^{31}_{15}$P		&$3d_{3/2}-2p_{1/2}$	&$-17.7(7.6)$	
&\cite{Aas:1982bb,Aas:1982vs}\\
			&$3d_{5/2}-2p_{3/2}$	&$0.4(2.6)$	
&\cite{Aas:1982bb,Aas:1982vs}\\
$^{40}_{20}$Ca		&$3d_{3/2}-2p_{1/2}$	&$-10(8)$	
&\cite{Hargrove:1977ue,PhysRevA.40.2176}\\
			&$3d_{5/2}-2p_{3/2}$	&$-3(6)$		
&\cite{Hargrove:1977ue,PhysRevA.40.2176}\\
$^{103}_{45}$Rh		&$4f_{5/2}-3d_{3/2}$	&$-3(28)$	
&\cite{Vuilleumier:1976pr,PhysRevA.40.2176}\\
			&$4f_{7/2}-3d_{5/2}$	&$18(27)$	
&\cite{Vuilleumier:1976pr,PhysRevA.40.2176}\\
$^{\rm nat}_{50}$Sn	&$4f_{5/2}-3d_{3/2}$	&$-6(7)$		
&\cite{Hargrove:1977ue,PhysRevA.40.2176}\\
			&$4f_{7/2}-3d_{5/2}$	&$-3(9)$		
&\cite{Hargrove:1977ue,PhysRevA.40.2176}\\
$^{\rm nat}_{56}$Ba	&$4f_{5/2}-3d_{3/2}$	&$0(7)$		
&\cite{Hargrove:1977ue}\\
			&			&$12(10)$	
&\cite{Tauscher:1977mw}\\
			&			&$-4(9)$		
&\cite{Dubler:1978rq}\\
			&$4f_{7/2}-3d_{5/2}$	&$-4(11)$	
&\cite{Hargrove:1977ue}\\
			&			&$17(9)$	
&\cite{Tauscher:1977mw}\\
			&			&$-12(9)$	
&\cite{Dubler:1978rq}\\
			&$5g_{7/2}-4f_{5/2}$	&$1(8)$	
&\cite{Hargrove:1977ue}\\
			&$5g_{9/2}-4f_{7/2}$	&$10(6)$	
&\cite{Hargrove:1977ue}\\
$^{\rm nat}_{58}$Ce	&$4f_{5/2}-3d_{3/2}$	&$1(10)$	
&\cite{Dubler:1978rq}\\
			&$4f_{7/2}-3d_{5/2}$	&$6(10)$	
&\cite{Dubler:1978rq}\\
$^{\rm nat}_{80}$Hg	&$5g_{7/2}-4f_{5/2}$	&$-32(29)$	
&\cite{Vuilleumier:1976pr}\\
			&$5g_{9/2}-4f_{7/2}$	&$-39(29)$	
&\cite{Vuilleumier:1976pr}\\
$^{203}_{81}$Tl		&$5g_{7/2}-4f_{5/2}$	&$-17(30)$	
&\cite{Vuilleumier:1976pr}\\
			&			&$-3(10)$	
&\cite{Dubler:1978rq}\\
			&$5g_{9/2}-4f_{7/2}$	&$-27(30)$	
&\cite{Vuilleumier:1976pr}\\
			&			&$-4(10)$	
&\cite{Dubler:1978rq}\\
			&			&$-10(7)$	
&\cite{Hargrove:1977ue}\\
$^{\rm nat}_{82}$Pb	&$5g_{7/2}-4f_{5/2}$	&$1(15)$	
&\cite{Hargrove:1977ue,PhysRevA.40.2176}\\
			&			&$0(13)$	
&\cite{Tauscher:1977mw,PhysRevA.40.2176}\\
			&			&$1(10)$	
&\cite{Dubler:1978rq,PhysRevA.40.2176}\\
			&$5g_{9/2}-4f_{7/2}$	&$-9(7)$	
&\cite{Hargrove:1977ue,PhysRevA.40.2176}\\
			&			&$23(12)$	
&\cite{Tauscher:1977mw,PhysRevA.40.2176}\\
			&			&$-6(10)$	
&\cite{Dubler:1978rq,PhysRevA.40.2176}\\
\hline\hline
 \end{tabular}
\end{center}
\end{table}

For most of the muonic transitions, the error from experiment and theory is roughly equal, and therefore reducing either could greatly improve these limits.  These experiments were all done during the 1970s and 1980s, therefore dramatic improvement in their measurement is possible.  On the theory side, 66\% of the error is from only two sources: electron screening and nuclear polarization~\cite{Borie:1982ax} which can also potentially be reduced.

\begin{figure}
\begin{center}
\includegraphics[width=\linewidth]{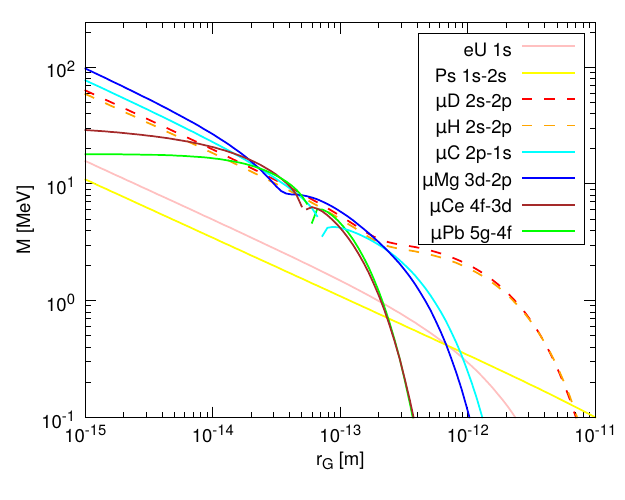}
\end{center}
\caption{\label{fig:alp0}Selected limits for $M$ as a function of $r_G$ with $\alpha=0$.  The solid lines correspond to $1\sigma$ lower bounds, while the dashed lines are the mean values of the discrepancies in muonic hydrogen and muonic deuterium.}
\end{figure}

To get a sense for the functional dependence of each transition on $r_G$, and $M$, in Fig.~\ref{fig:alp0} we have plotted a few example limits for the case $\alpha=0$.  The kinks appearing in the limits can be traced to the fact that the corrections in Table~\ref{tab:eqs} are positive semidefinite and negative semidefinite in different regions of $(r_G,M)$ space.  When $0<\alpha<1$, atoms with $m_i<m_p$ see their limits move higher, while for $m_i>m_p$ limits are weakened.  In this situation for example, the parameter space from $\mu$Mg is reduced while the positronium starts ruling out more space.  The tension between limits like this are responsible for a good deal of parameter space being unacceptable.  As will be seen, insisting that the $\mu^-p$ and $\mu^-D$ Lamb shifts are consistent place strong bounds on $\alpha$ 

\section{Analysis}   
\label{sec:anal}
We use the Bayesian inference tool \textsc{MultiNest} which calculates the evidence and explores parameter spaces with complex posteriors and pronounced degeneracies in high dimensions~\cite{Feroz:2007kg,Feroz:2008xx,2013arXiv1306.2144F}.  In addition to computing the evidence from the data, \textsc{MultiNest} derives the posterior probability distribution functions (PDFs) through application of Bayes' theorem.  As constraints, we take all the results in ~\Cref{tab:2sls,tab:1sls,tab:1s2s,tab:xray}.  We assume that the prior probability distribution function of each observable is given by a Gaussian with its standard deviation given by the uncertainty.  We have taken uniform logarithmic priors in $M=[10^{-5},10^5]$ MeV and $r_G=[10^{-18},10^{-10}]$ m and a uniform prior in $\alpha=[-3,3]$.

\begin{figure*}
\begin{center}
\includegraphics[width=\linewidth]{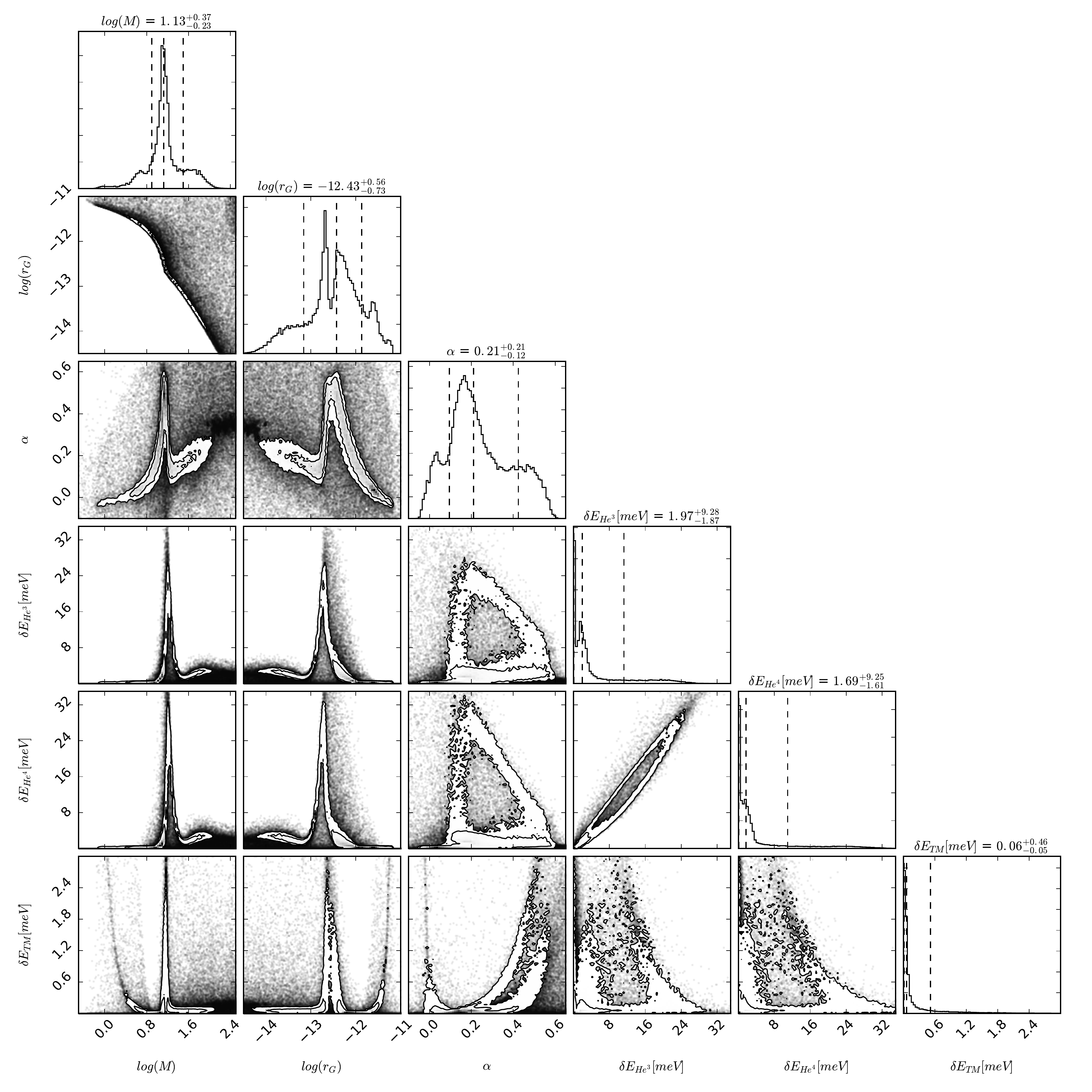}
\end{center}
\caption{\label{fig:parameters}1D and 2D marginal PDFs for $\log(M)$, 
$\log(r_G)$, and $\alpha$ produced using the Galileon contributions from 
Table~\ref{tab:eqs} to the transitions found in 
\Cref{tab:2sls,tab:1sls,tab:1s2s,tab:xray}. $M$ is in units of MeV and $r_G$ is in units of m.  Additionally plotted are the predictions for the $2s-2p$ 
Lamb shifts in $\mu^-He^3$, $\mu^-He^4$, and $\mu^-\mu^+$.  In the 1D plots, the dashed lines correspond to the mean, $1\sigma$ credible regions.  In the 2D plots, the contour regions are the $1\sigma$ and $2\sigma$ credible regions.}
\end{figure*}

While the full results of our calculation are found in Fig.~\ref{fig:parameters}, the mean values and $1\sigma$ credible intervals are $r_G=3.7^{+9.8}_{-3.0}\times10^{-13}$ m, $M=13^{+18}_{-6}$ MeV, and $\alpha=0.21^{+0.21}_{-0.12}$.  The mean value of $r_G$ found corresponds to a radius $\approx 425$ times larger than $r_p=0.8758(77)\times10^{-15}$ m.  This large value of $r_G$ is the same order of magnitude as the muonic hydrogen Bohr radius, implying that the orbitals themselves may be strongly modified.  The mean value of $M$ is excluded by LHC and astrophysical constraints, but these can be avoided by introducing chameleon interactions as stated above.  Our result for $M$ represents a limit, albeit model dependent, of $M>7$ MeV at the $1\sigma$ level. 

From the marginal PDFs, we see that a degeneracy exists between $r_G$ and $M$.  In contrast to~\cite{Lamm:2015gka} though the $2\sigma$ confidence region is finite and bounded.  In contrast, the value of $\alpha$ is restricted to a small range $\alpha\approx[0,0.6]$ because of heavy ions and leptonic systems.  The peak in $\alpha$ can be understood by considering the ratio of the energy correction to the $n\leq3$ transitions in two muonic atoms.  The ratio between two muonic systems $m_i>m_j$ is
\begin{equation}
  \frac{\delta E_{i}}{\delta E_{j}}\approx\left(\frac{m_{i}}{m_{j}}\right)^{1-4\alpha}\left(\frac{Z_i}{Z_j}\right)^3.
\end{equation}
Since increasing charge is related to increasing mass, the smallness of $\alpha$ prevents the mass-dependent term from dominating over the charge term except for very neutron-rich atoms, generically implying massive atoms have larger corrections.  In contrast, in the case of two isotopes, the charge term cancels. The ratio is then 
\begin{equation}
\label{eq:iso}
 \frac{\delta E_{i}}{\delta E_{j}}\approx\left(\frac{m_{i}}{m_{j}}\right)^{1-4\alpha}.
\end{equation}
Using this relation, we can see that for $\alpha>\frac{1}{4}$ heavier isotopes will have smaller corrections than lighter ones, and have larger corrections for $\alpha<\frac{1}{4}$.  If we insert the results from $\mu^- D$ and $\mu^- p$ into this relation, we see that they prefer a value of $\alpha=0.16$, which is near the peak of the 1D PDF of $\alpha$.  This indicates that the muonic Lamb shifts dominate the determination of $\alpha$.   

\section{Predictions for Helium and True Muonium}
\label{sec:hel}
Using the PDFs, it is possible to make predictions for the $2s-2p$ Lamb shift in $\mu^-He^3$ and $\mu^-He^4$ that will soon be presented by the CREMA Collaboration.  In Fig.~\ref{fig:parameters}, we present the PDFs for these two measurements and their relation to the model parameters.  We find the shifts to be $\delta E_{He^3}=1.97^{+9.28}_{-1.87}$ meV and $\delta E_{He^4}=1.69^{+9.25}_{-1.61}$ meV.  The mean value of these corrections is more than a factor of 4 larger than the discrepancies in muonic hydrogen and muonic deuterium, and are 0.1\% corrections to the theory values.  This would be easily measured by the CREMA Collaboration.  If a smaller value of $\Delta E_{He}$ is found, it has the ability to greatly restrict $(r_G,M,\alpha)$ space.

Additionally, the as-yet undiscovered bound state of true muonium $\mu^-\mu^+$ offers an opportunity to constrain the parameter space\cite{Jentschura:1997tv,Jentschura:1997ma,Karshenboim:1998am,Brodsky:2009gx,PhysRevD.91.073008,Lamm:2015fia,PhysRevA.94.032507}.  We can predict a correction to the Lamb shift of $\delta E_{TM}=0.06^{+0.46}_{-0.05}$ meV, which corresponds to a 0.1\% correction.  From Fig.~\ref{fig:parameters}, we see that the largest energy corrections in true muonium are in a different region of parameter space, and therefore are a strong complement to the muonic helium measurements.  Near-future experiments to detect and measure true muonium have been proposed~\cite{Celentano:2014wya,Banburski:2012tk,Benelli:2012bw,dirac,Nemenov:2001vp}.


As can be observed in Fig.~\ref{fig:parameters}, although the uncertainty on both predictions is large, they are strongly correlated.  The strong correlation between each muonic helium correction and the model parameters shows the upcoming measurements will have a large effect on restricting the entire $(r_G,M,\alpha)$ parameter space.  From the insensitivity of Eq.~(\ref{eq:iso}) to $(r_G,M)$, combining both muonic helium measurements is greater than merely the sum of their parts. 
\section{Summary and Conclusions}
\label{sec:sum}
In this paper, we have shown that Galileon corrections to muonic hydrogen and muonic deuterium can be consistently explained by introducing a one-parameter family of relationships between the cutoff radii of different systems.  Furthermore, predictions for the corrections to upcoming muonic helium experiments have been made.  These corrections are can be quite large and the CREMA Collaboration's upcoming results will dramatically reduce the parameter space.

In the future, other than improving the experimental and theoretical errors of the current measurements, another important direction to investigate would be computing the corrections to other observables.  Computing the fine and hyperfine splittings due to the Galileon couplings would be useful given there are no discrepancies in these measurements.  A very fruitful direction of study would be in the calculation of the corrections to the anomalous magnetic moment of leptons, ($a_\ell$).  Combining the high precision measurement of $a_e$ with the persisting anomaly in $a_\mu$ would be useful in restricting the parameter space of $(r_G,M,\alpha)$. 

\begin{acknowledgments}
HL would like to express his gratitude to Jayden Newstead for his insightful comments.  This work was supported by the National Science Foundation under Grants No. 
PHY-1068286 and No. PHY-1403891.
\end{acknowledgments}
\bibliographystyle{apsrev4-1}
\bibliography{/home/hlamm/wise}
\end{document}